\documentstyle[12pt]{article}

\begin{document}
\renewcommand{\thefootnote}{\fnsymbol{footnote}}
\begin{flushright}
KOBE--FHD--97--01\\
March~~~~~~~~1997
\end{flushright}
\begin{center}
{\LARGE\bf Semi--inclusive {\boldmath $\Lambda_c^+$} Leptoproductions and\\
\vspace{0.5em}
Polarized Gluon Distributions}\\

\vspace{3.5em}

N. I. Kochelev\footnote[2]{E--mail~~~~kochelev@thsun1.jinr.dubna.su}\\
\vspace{0.8em}
{\it Bogoliubov Laboratory of Theoretical Physics}\\
{\it Joint Institute for Nuclear Research}\\
{\it 141980, Dubna, Moscow region, Russia}\\
\vspace{1em}
T. Morii\footnote[3]{E--mail~~~~morii@kobe--u.ac.jp}\\
\vspace{0.8em}
{\it Faculty of Human Development, Division of}\\
{\it Sciences for Natural Environment}\\
{\it and}\\
{\it Graduate School of Science and Technology,}\\
{\it Kobe University, Nada, Kobe 657, Japan}\\
\vspace{1em}
and\\
\vspace{1em}
T. Yamanishi\footnote[8]{E--mail~~~~yamanisi@natura.h.kobe--u.ac.jp}\\
\vspace{0.8em}
{\it Research center for Nuclear Physics,}\\
{\it Osaka University, Ibaraki, Osaka 567, Japan}\\
\vspace{4.5em}
{\bf Abstract}
\end{center}
In order to extract the behavior and magnitude of the spin--dependent gluon
distribution, we propose a semi--inclusive $\Lambda_c^+$ production using
unpolarized lepton beams and polarized proton targets.
The correlation between the target proton spin and the spin of $\Lambda_c^+$ 
produced in the target fragmentation region might be very
effective for testing various models of polarized gluons.

\baselineskip=24pt

\vfill\eject

The analysis of abundant experimental data on deep inelastic
scatterings at large momentum transfer squared $Q^2$ indicates that
a proton has structure and is composed of constituents called
partons which are identified as massless quarks and gluons.
The quantum number of the proton is carried by those constituents.
For instance, the proton spin is given by the sum of the spin of
those constituents and their orbital angular momenta,
\begin{equation}
\frac{1}{2}=\frac{1}{2}\Delta\Sigma+\Delta g+\langle L_Z\rangle_{q+g}~,
\label{eqn:sumrule}
\end{equation}
where $\Delta\Sigma$ and $\Delta g$ are the amount of the proton spin carried
by quarks and gluons, respectevely, and
$\langle L_Z\rangle_{q+g}$ implies the orbital angular momenta of quarks
and gluons.

Recent measurement of spin--dependent structure functions of nucleons
$g_1(x, Q^2)$, carried by EMC and SMC at CERN and E142, E143 and E154
at SLAC for
polarized deep inelastic scatterings (DIS), has been widely interpreted as
evidence of a breakdown of the simple quark--parton model of nucleons,
with a surprizing result that a little of the proton spin is
carried by quarks, $\Delta\Sigma\sim 0.30$\cite{pDIS}.
Several competing explanations for this result have been proposed so far.
Among them, there exist two most common interpretations:
either the strange sea--quark (s--quark) of the proton is significantly
polarized, or gluons contribute to the spin--dependent proton structure
function through a flavor singlet current with subtle anomaly effects.
In order to understand the underlying dynamics of
the spin sum rule of eq.(\ref{eqn:sumrule}), it is
important to know the behavior of polarized s--quarks and gluons
in a proton.
Many people have proposed various polarized parton
distributions (PPDs) from a fit to recent data on $g_1(x,Q^2)$ 
based on above interpretations\cite{GS95,BBS95,GRV95}. From
these analyses, one can see that the polarized distribution function of
gluons can be largely different among various models as shown in fig.1,
whereas those models mostly reproduce the valence quark polarization
in a nucleon as well\cite{Gehrmann95}.
The values of $\Delta g$ are large or small, depending on models.
Far from that, some people have discussed about the possiblility of even
negatively polarized gluons\cite{Jaffe96,Kochelev96}.
In any case, at present the behavior and magnitude of the gluon polarization
in a proton is still obscure.
In order to understand the nucleon spin structure, their experimental
determination is urgent.
However, from measurement of $g_1(x, Q^2)$ alone, one
cannot constrain the behavior of polarized gluons and test their models.
Even if $g_1(x, Q^2)$ is measured with much higher precision, the
situation cannot be improved as long as we
remain in the polarized inclusive processes,
$\vec\ell \vec p\to\ell^{'} X$, alone.
Therefore, we are forced to proceed to other processes such as 
polarized proton--polarized
proton collisions or semi--inclusive lepton--nucleon collisions
for polarized DIS, $\vec\ell \vec p\to\ell^{'} h X$.

On the other hand, a large acceptance and high 
luminosity measurement of identified
hadrons in DIS of polarized muons on polarized solid targets is now
proposed by COMPASS Collaboration at CERN\cite{COMPASS}.
One of purposes of COMPASS experiments 
is to extract the spin--dependent
distribution of individual partons from these semi--inclusive processes,
in particular, by analyzing charmed hadron productions
whose dominant processes are photon--gluon fusion.  
In the scattering of polarized lepton beams on polarized proton targets,
photon--gluon fusion for heavy quark productions such as
J/$\psi$ productions, open charm productions and
so on, is generally a good process for testing gluon
polarizations and many people have discussed 
interesting analyses so far\cite{charm}.

In this letter, we propose a different process to test 
the gluon polarization in a proton,
which is a semi--inclusive production of polarized $\Lambda_c^+$
baryons from a polarized proton, $\ell \vec p\to \ell^{'} \vec \Lambda_c^+ X$,
using unpolarized lepton beams.
Since a process of a charm quark decaying into  $\Lambda_c^+$ almost carries
the semi--inclusive production of $\Lambda_c^+$ and 
a contribution of photon--light quark processes is smaller than
the case of open charm productions, the ambiguity
of this process might be less than the case of open charm
productions\cite{Kaidalov86}.
In addition, determining the polarization of $\Lambda_c^+$ is
considerably easier because $\Lambda_c^+$ works as a self--analyzing
particle.
As is well known, the spin of $\Lambda_c^+$ is carried by a charm quark.
Since a charm is created by a gluon via photon--gluon fusion 
as shown in fig.2 and thus has a spin parallel to
the gluon spin, the direction of the $\Lambda_c^+$ spin which preseves
the charm quark spin, 
depends on that of the gluon spin.  Therefore, 
if gluons are largely positively--polarized in a nucleon, 
the correlation between the target proton spin and 
the spin of $\Lambda_c^+$ produced in the
target fragmentation region $x_{_F}<0$ along the target polarization axis 
is expected to become positive because the process
can occur only via photon--gluon fusion in the lowest order.
This suggests that the spin correlation between the target proton and
produced $\Lambda_c^+$ can give a good
information on gluon polarization in a proton.

Now, we define a spin correlation asymmetry $A_{LL}$ as an interesting
observable parameter,
\begin{eqnarray}
A_{LL}~&=&~\frac{\left[d\sigma_{++}-d\sigma_{+-}+
d\sigma_{--}-d\sigma_{-+}\right]}
{\left[d\sigma_{++}+d\sigma_{+-}+
d\sigma_{--}+d\sigma_{-+}\right]}
\nonumber\\
&=&~\frac{d\Delta\sigma/dy}{d\sigma/dy}~,
\label{eqn:A_LL}
\end{eqnarray}
where $d\sigma_{+-}$, for instance, denotes that the spin
of the target proton and produced $\Lambda_c^+$ is positive and
negative, respectively.
The spin--dependent differential cross section for
$\gamma^* \vec p\to\vec \Lambda_c^+ X$ which is related to that
for $\ell \vec p\to \ell^{'} \vec \Lambda_c^+ X$ is given by
\begin{eqnarray}
\frac{d\Delta\sigma^{\gamma^* p\to\Lambda_c^+ X}}{dy}(s, Q^2) &=&
\int^{p_{T max}^2}_{p_{T min}^2}\int^1_{x_{min}}dp_T^2dx~\Delta g(x, Q^2)~
\Delta D_{\Lambda_c^+/c}(z)~f(s, Q^2)\nonumber\\
&&\times\frac{d\Delta\hat \sigma^{\gamma^* \vec g\to\vec c~\bar c}}{d\hat t}
(\hat s,\hat t,\hat u)~,
\label{eqn:dDs/dy}
\end{eqnarray}
with
\begin{equation}
f(s, Q^2) = \frac{x \sqrt s (s+Q^2)}
{(M_{\Lambda_c^+}^2+p_T^2)^{\frac{1}{2}}[se^y+\{xs-Q^2(1-x)\}e^{-y}]}~,
\label{eqn:f-jacb}
\end{equation}
where $\Delta g(x, Q^2)$ and $\Delta D_{\Lambda_c^+/c}(z)$ represent
the polarized gluon distribution and spin--dependent fragmentation
function of an outgoing charm quark decaying into a polarized $\Lambda_c^+$
with momentum fraction $z$, respectively.
Unfortunately, at present the spin--dependent fragmentation function
$\Delta D_{\Lambda_c^+/c}(z)$ is not known for lack of
experimental data though we have some knowledge of the spin--independent one
$D_{\Lambda_c^+/c}(z)$ \cite{DeGrand,Peterson}.
However, since a charm quark is heavy, it might not be unreasonable
to use $D_{\Lambda_c^+/c}(z)$ for
$\Delta D_{\Lambda_c^+/c}(z)$.
Here we use $D_{\Lambda_c^+/c}(z)$ suggested in
ref.\cite{Peterson} for $\Delta D_{\Lambda_c^+/c}(z)$.
The spin--independent cross section is given by just replacing
$\Delta g(x, Q^2)$ and $d\Delta\hat\sigma/d\hat t(\hat s,\hat t,\hat u)$
in eq.(\ref{eqn:dDs/dy}) by $g(x, Q^2)$ and
$d\hat\sigma/d\hat t(\hat s,\hat t,\hat u)$, respectively.
The perturbative QCD allows to calculate the subprocess
cross section of photon--gluon fusion,
$\gamma^* \vec g\to\vec c~\bar c$.
Since the spin--independent subprocess cross section has been already
presented in ref.\cite{Fontannaz},
we here show only the spin--dependent one
\begin{eqnarray}
&&\frac{d\Delta\hat \sigma}{d\hat t}(\hat s,\hat t,\hat u)=
\frac{\pi\alpha_s\alpha e_q^2}{2\hat s^2}
\left[-\frac{\hat s(A+B)}{(\hat s+Q^2)}\left(\frac{\hat u_1}{\hat t_1}+
\frac{\hat t_1}{\hat u_1}+2\right)\right.
\label{eqn:dDs/dt}\\
& &\left. +(A-B)(\hat s+Q^2)\left\{\frac{\hat s+\hat u_1+2m_c^2}{\hat t_1^2}+
\frac{2m_c^2-\hat u_1}{\hat u_1^2}+
\frac{2(2m_c^2-\hat s)}{\hat t_1\hat u_1}\right\}\right]~,
\nonumber
\end{eqnarray}
with
\begin{eqnarray}
& &\hat t_1 = \hat t-m_c^2~,~~\hat u_1 = \hat u-m_c^2~,
\label{eqn:AB}\\
& &A=\cos\theta_c\approx\cos\theta_{\Lambda_c^+}=
\frac{(M_{\Lambda_c^+}^2+p_T^2)^{\frac{1}{2}}(e^y-e^{-y})}
{\left\{M_{\Lambda_c^+}^2(e^y-e^{-y})^2+p_T^2(e^y+e^{-y})^2
\right\}^{\frac{1}{2}}}~,~~
B=\left(1-\frac{4m_c^2}{\hat s}\right)^{\frac{1}{2}}~.\nonumber
\end{eqnarray}
Here we take approximately $\theta_c\approx\theta_{\Lambda_c^+}$ for the
scattering angle of the charm quark and that of produced $\Lambda_c^+$, which
can be expressed in terms of the transverse momentum and rapidity of
produced $\Lambda_c^+$.

In order to examine how the observed parameter is sensitive to
the behavior of polarized gluon distributions, we calculate $A_{LL}$ by
using four typical examples of $\Delta g(x, Q^2)$ presented in fig.1.
At a CMS energy of the virtual photon--proton collision, $\sqrt s=10$GeV
(which corresponds to $\gamma^*$ energy $\nu=56$GeV)
and a momentum transfer squared
$Q^2=10$GeV$^2$, whose kinematical region can be covered by COMPASS
experiments, we have calculated the asymmetry $A_{LL}$ and shown it in
fig.3 as a function of rapidity $y$ of $\Lambda_c^+$.
In figs.3 and 4, to calculate unpolarized cross sections, we have used
an unpolarized gluon distribution of ref.\cite{Owens} 
for the solid and dotted lines, the one of
ref.\cite{BBS95} for the dash-dotted line and the one of ref.\cite{GRV92} 
for the dashed line.
One can see that the $A_{LL}$ significantly depends on the behavior and 
magnitude of polarized gluon distributons.
It is remarkable that although the lower limit of 
the integration variable $x$ for the
differential cross section in eq.(\ref{eqn:dDs/dy}) is $x_{min}=0.29$
for $\sqrt s=10$GeV and $Q^2=10$GeV$^2$ at $y=-0.5$ and 
the difference of $\Delta g(x)$ among various models
taken up here is not so big for the region larger than $x\approx 0.3$,
the difference of $A_{LL}$ becomes conspicuous.
This implies that the process might be very promising to 
distinguish various models of polarized gluons.

Some comments are in order.  (1) In order to determine the polarization of
$\Lambda_c^+$ in experiment, one must observe the processes 
of $\Lambda_c^+\to\Lambda\pi^+$
or $\Lambda e^+\nu_e$, whose branching ratios are
(7.9$\pm$1.8)$\times$10$^{-3}$ or 1.4$\pm$0.5\%, respectively.
Since the spin--dependent and spin--independent cross sections 
are not so large, being an order of 10$^{-2}$nb 
and 10$^{-1}$nb, respectively, at
$\sqrt s=10$GeV and $Q^2=10$GeV$^2$ as shown in fig.4,
we need rather high luminosity, which 
can be hopefully obtained.
(2) For the semi--inclusive $\Lambda_c^+$ production, there might be other
processes in addition to the process calculated here:
a vector meson dominance process, a diffractive dissociation and
a compton scattering process due to an intrinsic charm component.
However, by constraining a kinematical region as $x_{_F}<0$ for 
produced $\Lambda_c^+$, we can pick up the photon--gluon fusion mechanism
which dominantly contributes to
$\ell \vec p\to \ell^{'} \vec \Lambda_c^+ X$.
In such a region, the value of spin correlations between the target proton
and produced $\Lambda_c^+$ can give a good information on gluon
polarizations in a proton.
(3) Some people have discussed that
measurement of polarization of $\Lambda$ produced in
the target fragmentation region for unpolarized lepton scatterings
off a polarized proton target could determine whether s--quarks 
polarized negatively or gluons polarized positively 
in a proton\cite{Ellis}.
In this reaction, one can expect that
for the latter case the correlation between the target proton spin and 
the spin of $\Lambda$ produced along the target polarization axis becomes
positive, while that should be negative for the former case.
Our prediction is deeply related to their prediction.  Experimental
test of both predictions would lead us to a good understanding
of the spin structure of a proton.

In summary, we have calculated the spin correlation
asymmetry $A_{LL}$ for the process with unpolarized lepton beams and 
polarized proton targets, $\ell \vec p\to \ell^{'} \vec \Lambda_c^+ X$,
and found that the behavour of $A_{LL}$ is sensitive to
the form of $\Delta g$.  We hope the present prediction can be tested
in the forthcoming experiment.

One of the authors(T.M.) would like to thank Prof. V. Burov for his kind 
hospitality during his visit to Dubna in 1996.  N. K. would like to thank 
the JSPS and the 90th anniversary committee of Kobe university for their 
financial support for developing the collaboration at Kobe.

\vfill\eject

\begin{center}
{\large \bf Figure captions}
\end{center}
\begin{description}
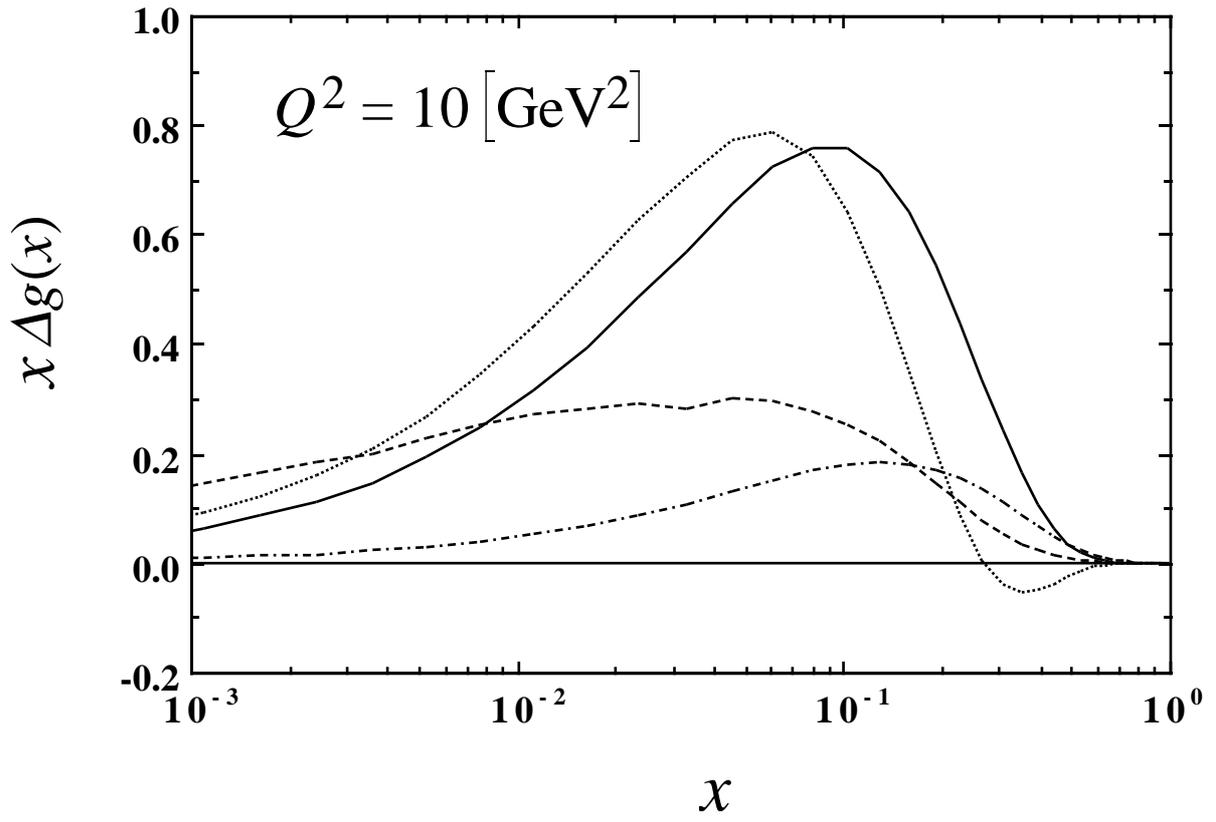
\item[Fig. 1:] The $x$--dependence of polarized gluon distributions
at $Q^2=10$GeV$^2$.
The solid, dotted, dash--dotted and dashed lines indicate the set A, C
of ref.\cite{GS95}, ref.\cite{BBS95} and the `standard scenario' of
ref.\cite{GRV95}, respectively.

\vspace{2em}

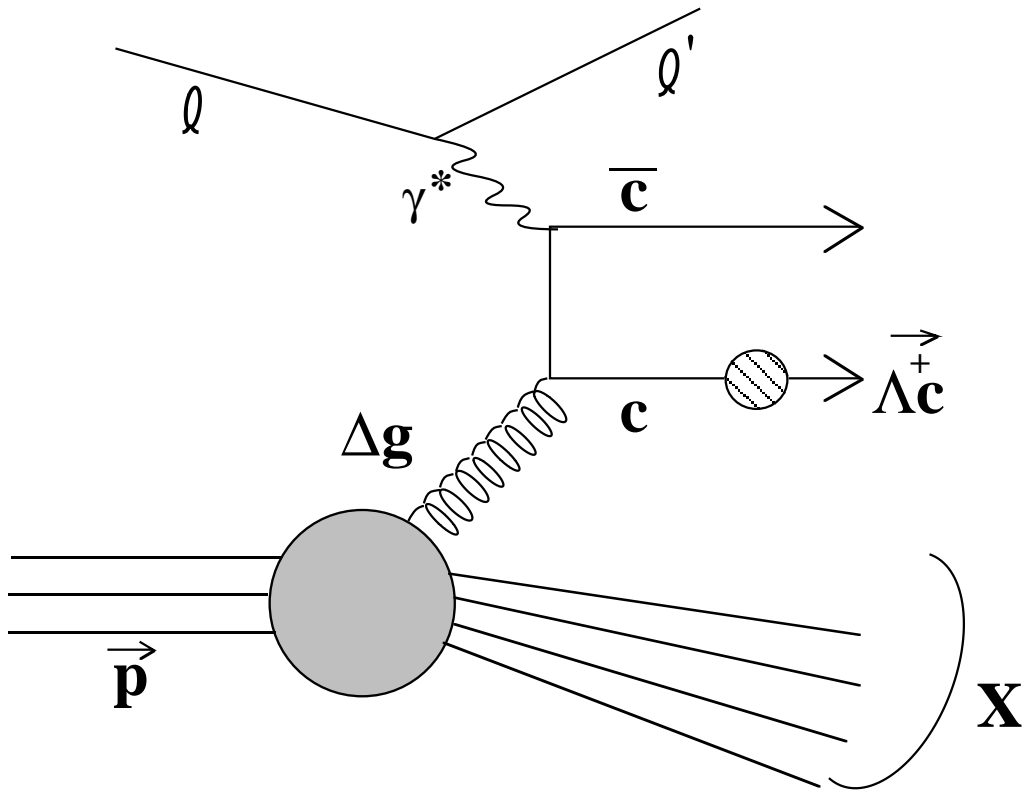
\item[Fig. 2:] The lowest order QCD diagram for $\Lambda_c^+$
leptoproductions in unpolarized lepton--polarized proton scatterings.

\vspace{2em}

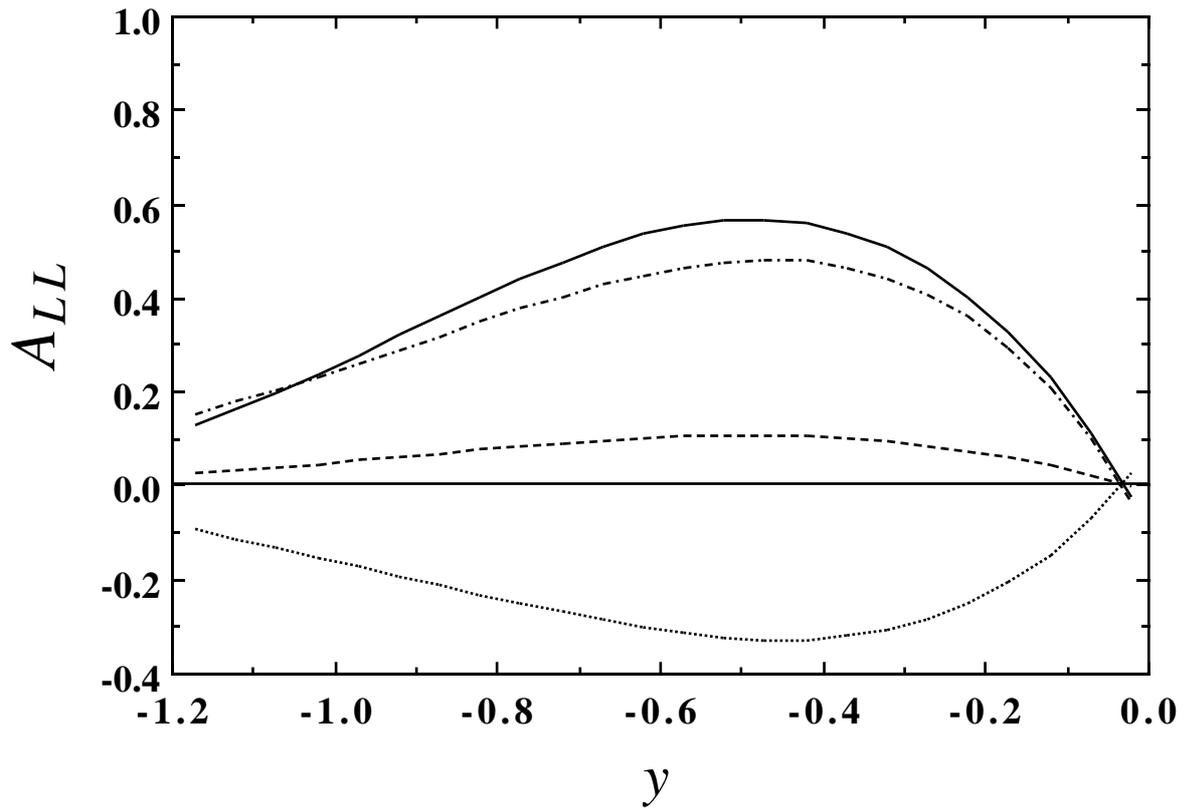
\item[Fig. 3:] The spin correlation asymmetry $A_{LL}$ as a function of
rapidty $y$ at $\sqrt s=10$GeV and $Q^2=10$GeV$^2$.
Various lines represent the same as in fig.1.

\vspace{2em}

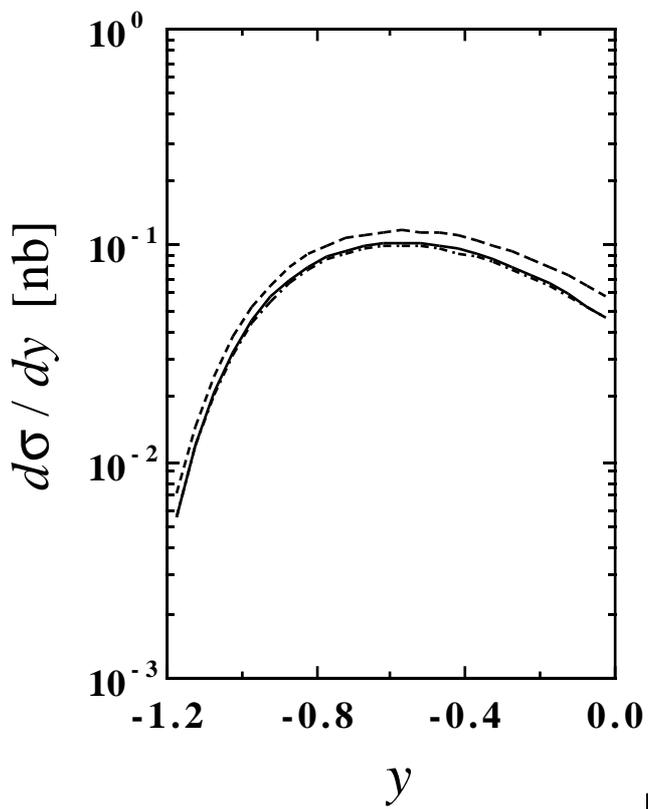
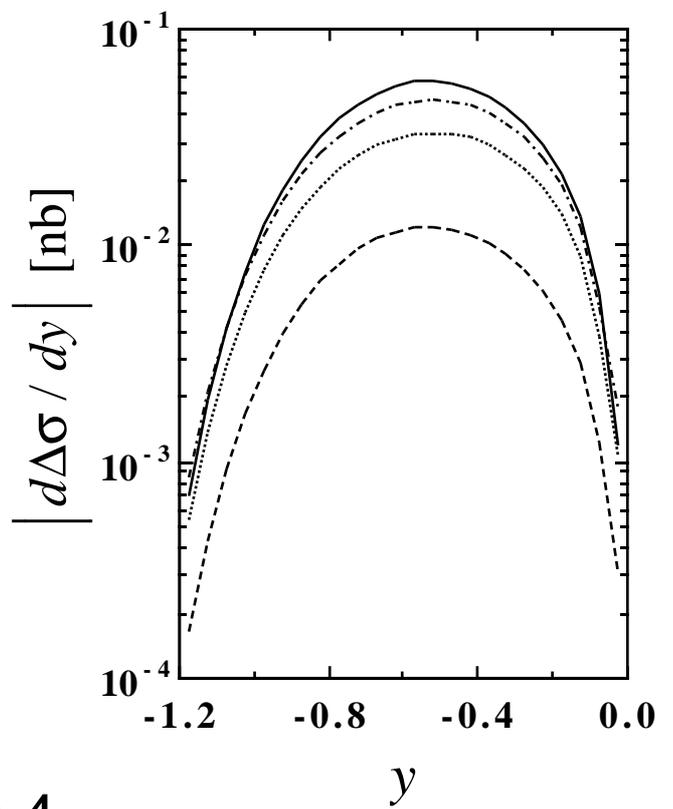
\item[Fig. 4:] The spin--dependent and spin--independent differential
cross sections as a function of $y$ at $\sqrt s=10$GeV and $Q^2=10$GeV$^2$.
Various lines represent the same as in fig.1.
\end{description}
\end{document}